 \documentclass[11pt,english]{article}

\usepackage{graphicx,psfrag,epsfig,amsfonts,amssymb,amsmath,babel}

 \usepackage[colorlinks=true]{hyperref}

\hypersetup{urlcolor=blue, citecolor=red}

\newtheorem{definition}{Definition}
\newtheorem{remark}[definition]{Remark}

\newtheorem{theorem}[definition]{Theorem}
\newtheorem{proposition}[definition]{Proposition}

\begin{document}

\title{Dale's Principle is necessary for \\ an optimal neuronal network's dynamics}

\author{Eleonora Catsigeras\thanks{Instituto de Matem\'{a}tica y Estad\'{\i}stica \lq\lq Prof. Ing. Rafael Laguardia\rq\rq \  (IMERL),
 Fac. Ingenier\'{\i}a,  Universidad de la Rep\'{u}blica,  Uruguay.
 E-mail: eleonora@fing.edu.uy
  EC was partially supported by CSIC of Universidad de la Rep\'{u}blica and ANII of Uruguay.} }

\date{}

\maketitle

\noindent The final revised version of this preprint will be published in  Applied Mathematics (Irvine), ISSN 2152-7385,  Special Issue on Biomathematics  to appear at

\hfill  www.scirp.org/journal/am/

\begin{abstract}
We study a mathematical model of biological neuronal networks composed by any finite number $N \geq 2$ of non necessarily identical cells. The model is a deterministic dynamical system governed by finite-dimensional impulsive differential equations.  The statical structure of the network is described by a directed and weighted graph whose nodes are certain subsets of neurons, and whose edges are the groups of synaptical connections among those subsets. First, we prove that among all the possible networks  such that their respective graphs are  mutually isomorphic, there exists a dynamical optimum. This optimal network exhibits the richest dynamics: namely, it is capable to show the most diverse set of responses (i.e. orbits in the future) under external stimulus or signals.   Second, we prove that all the neurons of a dynamically optimal   neuronal network  necessarily satisfy Dale's Principle, i.e. each neuron must be either excitatory or inhibitory, but not mixed. So, Dale's Principle is a mathematical necessary consequence of a theoretic optimization process of  the dynamics of the network.  Finally, we prove that Dale's Principle is not sufficient for the dynamical optimization of the network.

\end{abstract}

%%%%%%%%%%%%%%%%%%%%%%%

 {\noindent   {{\em MSC }2010:  {\  Primary: 92B20. Secondary: 37N25, 34A37, 05C22, 05C79.} }}

\section{Introduction}
\label{SectionIntroduction} \vspace{-4pt}

 Based on experimental evidence, Dale's Principle in Neuroscience (see for instance \cite{1,2}) postulates   that   most neurons of a biological  neuronal network   send  the same set of biochemical substances (called neurotransmitters)  to   the   other neurons that are connected with them. Most neurons release more than one neurotransmitter,  which is called the \lq\lq co-transmission" phenomenon \cite{7,8}, but \em the set \em of neurotransmitters is constant for each  cell. Nevertheless, during plastic phases of the nervous systems, the neurotransmitters that are released by certain groups of neurons change according to the development of the neuronal network. This plasticity allows the network   perform   diverse and adequate dynamical responses   to external stimulus: \lq\lq Evidence suggests that
during both development
(in utero) and the postnatal
period, the neurotransmitter
phenotype of neurons is plastic
and can be adapted as a function
of activity or various
environmental signals\rq\rq \ \cite{8}. \ Also a
certain
phenotypic plasticity
occurs   in some cells of the nervous system of mature animals, \lq\lq
suggesting that a dormant
phenotype can be put in play
by external inputs\rq\rq \ \cite{8}.

 Some mathematical models of the neuronal networks represent them as deterministic dynamical systems (see for instance \cite{MirolloStrogatzPulseCoupledNetworks1990,BudelliLeakyIntegratorPacemakerModel,
 MassBishop2001BookPulsedNeuralNetworks,ErmentroutDynamicsModelsExcitability}). In particular, the dynamical evolution of the state of each neuron during the interspike intervals, and the dynamics of the bursting phenomenon, can be modelled  by a finite-dimensional ordinary differential equation (see for instance \cite{ErmentroutDynamicsModelsExcitability, SpikingNeuronModels} and in particular \cite{Coombes2013} for a   mathematical model of a neuron as a dynamical system evolving on a multi-dimensional space). When considering a network of many neurons, the   synaptical connections are frequently modelled by impulsive coupling terms between the   equations of the many cells (see for instance \cite{MassBishop2001BookPulsedNeuralNetworks,Izhikevich_PulseCoupledDynamicsExcitabilityBursting, ImpulsiveDifferentialEquationsSpikingSynapsesModels,CatGuiraud}). In such a mathematical   model,  Dale's Principle is translated into the following statement:

{\bf Dale's Principle: } \em Each neuron is either inhibitory or excitatory. \em We recall that a neuron $i$ is called inhibitory (resp. excitatory) if  its spikes  produce, through the electro-biochemical actions that are transmitted along the axons of $i$, only null or negative (resp. positive)   changes in the membrane potentials of all the other neurons $j \neq i$ of the network. The amplitudes of those changes may depend on many  variables. For instance, they may depend on the membrane instantaneous permeability of the receiving cell $j$.   But the sign of the postsynaptical actions is  usually only   attributed   to the electro-chemical properties of the  substances that are released by the sending   cell $i$. In other words, the sign   depends only on the set of neurotransmitters that are released by $i$. Since this set of  substances is   fixed for each neuron $i$ (if $i$ satisfies Dale's Principle), the sign of its synaptical actions on the other neurons $j \neq i$ is fixed for each   cell $i$, and thus, independent of the receiving neuron $j$.

In this paper we adopt  a simplified mathematical model of the neuronal network with a finite number $N \geq 2$ of neurons, by means of a system of deterministic impulsive differential equations. This model is taken from \cite{Izhikevich_PulseCoupledDynamicsExcitabilityBursting, CatGuiraud}, with an adaptation that allows   the state variable $x_i$ of each cell $i$ be multidimensional. Precisely $x_i$ is a vector of finite dimension, or equivalently, a point in a finite-dimensional manifold of an Euclidean space. The finite dimension of the state variable $x_i$ is larger or equal than 1, and besides, may depend on the neuron $i$. The dynamical model of the network is the solution of a system of impulsive differential equations. This dynamics evolves on a product manifold whose dimension is the sum of the dimensions of the state variables of its $N$ neurons.

We do not assume  a priori that the neurons of the network  satisfy Dale's Principle. In Theorem \ref{theoremDalePrinciple} we prove  this principle a posteriori, as a necessary final consequence of  a dynamical optimization process. We assume that during this process, a plastic phase of the neuronal network occurs, eventually changing the total numbers of neurons and synaptical connections, but such that the graph-scheme of the synaptic connections among \em groups of mutually identical cells \em remains unchanged. We assume that a maximal amount of dynamical richness is pursued during such a plastic development of the network. Then, by means of a rigourous deduction  from the abstract  mathematical model, we prove that, among all   the mathematically theoretic networks ${\mathcal N}$ of such a model, \em those   exhibiting an optimal dynamics (i.e. the richest or the most versatile dynamics)   necessarily satisfy
Dale's Principle \em (Theorem \ref{theoremDalePrinciple}).

The mathematical criteria to decide the dynamical optimization is the following:  First, in Definition \ref{definitionEquivalentNetworks}, we classify all the theoretic  neuronal networks (also those that hypothetically do not satisfy Dale's Principle)   into non countably many equivalence classes. Each class  is a family of mutually equivalent networks,  with respect to their internal synaptical connections among groups of cells (we call those groups of cells \em synaptical units \em  in Definition \ref{definitionSynapticalUnit}).    Second, in Definitions \ref{definitionRicherNetwork} and \ref{definitionDynamicalOptimum}, we agree to say that a network ${\mathcal N}$ has an \em optimal dynamics   conditioned to its class, \em if  the dynamical system modelling  any other network ${\mathcal N}'$ in the same class as ${\mathcal N}$,  has a space of  orbits in the future that \em is  a subset \em of the space  of orbits of    ${\mathcal N}$. In other words, ${\mathcal N}$ is the network  capable to perform the   richest dynamics, namely, the most diverse set of possible evolutions in the future   among all the networks that are in the same class.

\vspace{.3cm}

\noindent{\bf RESULTS TO BE PROVED: }

\noindent In {\bf Theorem \ref{theoremExistenceofOptimalNetworks}}  we prove that \em the theoretic dynamical   optimum exists \em in any equivalence class of networks that have isomorphic synaptical graphs.

\noindent  In {\bf Main Theorem \ref{theoremDalePrinciple}} we prove that  such an \em   optimum is achieved   only if the network ${\mathcal N}$ satisfies Dale's Principle. \em

\noindent  In {\bf Main Theorem \ref{theoremCounterexample}} we prove that the converse of Theorem \ref{theoremDalePrinciple} is false: \em Dale's Principle is not sufficient for a network  exhibit the optimal dynamics \em within its synaptical equivalence class.

The results are abstract and theoretically deduced from the mathematical model. They are epistemologically suggestive since they give  a possible answer to the following question:

\noindent {\bf Epistemological question: } \em Why \em does Dale's Principle  hold  for most cells in the nervous systems of animals?

Mathematically, the hypothesis of searching  for an optimal dynamics   implies (through Theorem \ref{theoremDalePrinciple}) that at some step of the optimization process all the cells must  satisfy Dale's Principle. In other words, this principle would be a consequence, instead of a cause, of   an optimization process during the plastic phase of the network. This conclusion   holds under the hypothesis that   the dynamical optimization (i.e. the maximum  dynamical richness) is one of the \lq\lq natural\rq\rq \ pursued aims during a certain changeable development   of the network.

Finally, we notice that the converse of Theorem \ref{theoremDalePrinciple} is false:  there exist mathematical examples of simple abstract  networks whose cells satisfy Dale's Principle but are not dynamically optimal (Theorem \ref{theoremCounterexample}). Thus, Dale's Principle is necessary but not sufficient for the dynamical optimization of the network.

\vspace{.2cm}

\noindent{\bf Structure of the paper and purpose of each section:}

 In Section \ref{sectionModel} we write the hypothesis of Main Theorems \ref{theoremDalePrinciple} and \ref{theoremCounterexample}.
From  Section  \ref{sectionGraph&Layers} to \ref{sectionDalePrinciple}  we prove  Main Theorem \ref{theoremDalePrinciple}. The proof is developed  in  four steps, one in each separate section. The first step (Section \ref{sectionGraph&Layers})   is devoted to prove the intermediate result of Proposition \ref{propositionUnits}.  The second step (Section \ref{sectionEquivalentNetworks}) is deduced from   Proposition \ref{propositionUnits}. The third step (Section \ref{sectionDynamicalOptimum}) is logically independent from the first and second steps, and is devoted to obtain the two intermediate results of Proposition \ref{propositionNumberOfNeuronsLarger} and Theorem \ref{theoremExistenceofOptimalNetworks}. Section \ref{sectionDalePrinciple} exposes the fourth step (the end) of the proof of  Main Theorem \ref{theoremDalePrinciple},  from the logic junction  of the   previous three steps,   using the intermediate results (Propositions \ref{propositionUnits} and \ref{propositionNumberOfNeuronsLarger}, and Theorem \ref{theoremExistenceofOptimalNetworks}).

On  the one hand, the   intermediate results (Propositions \ref{propositionUnits} and \ref{propositionNumberOfNeuronsLarger}, and Theorem \ref{theoremExistenceofOptimalNetworks}) are \em necessary stages \em in the logical process of our proof of  Main Theorem \ref{theoremDalePrinciple}  which ends in Section \ref{sectionDalePrinciple}. In fact,   Theorem \ref{theoremDalePrinciple}  establishes that, \em if there  exists \em a dynamical optimum within each synaptical equivalence class of networks, this optimal network   necessarily satisfies Dale's Principle. But this result would be void if we did not prove (as an intermediate step), that a   dynamically optimal network exists (Theorem \ref{theoremExistenceofOptimalNetworks}). It would be also   void if we did not prove that the synaptical equivalence classes of networks exist (Definition \ref{definitionEquivalentNetworks}). The synaptical equivalence classes of networks could not been defined if the inter-units graph of the network did not existed (Definition \ref{definitionInter-UnitsGraph}). And  these graph exists as an immediate corollary of Proposition \ref{propositionUnits}. So, Proposition \ref{propositionUnits} must be proved   as an intermediate step for our final purpose. Finally, the end of the proof of Main Theorem \ref{theoremDalePrinciple} argues by contradiction: if the dynamically optimal network did not satisfy Dale's Principle, then Proposition  \ref{propositionNumberOfNeuronsLarger} would be false. So, we need first, also as an intermediate step, to prove Proposition \ref{propositionNumberOfNeuronsLarger}.

On the other hand, to prove all the required   intermediate results, we need   some other  (previous) mathematical statements from which we deduce the intermediate results. So, we start   posing  all the  previous mathematical statements (obtaining them from the general hypothesis of Section \ref{sectionModel}), in a series of  mathematical definitions, comments and remarks that are at the beginning of  Sections \ref{sectionGraph&Layers}, \ref{sectionEquivalentNetworks} and \ref{sectionDynamicalOptimum}.

In Section \ref{sectionCounterexample}, we end the proof of   Main Theorem \ref{theoremCounterexample} stating that Dale's Principle is not sufficient for the dynamical optimization. Its final statement is proved by applying directly some of the definitions, intermediate results and examples of Sections \ref{sectionGraph&Layers},  \ref{sectionEquivalentNetworks}  and \ref{sectionDynamicalOptimum} (in particular, those of Figures \ref{Figure1}, \ref{Figure2} and \ref{Figure3}).

Finally, in Section \ref{sectionConclusions} we write the   conclusions obtained from all the mathematical  results that are proved along the paper.

 \vspace{-.5cm}

\section{\bf The   hypothesis  \\ (The model by a system of impulsive differential equations)} \label{sectionModel}

\vspace{-.3cm}

We assume a simplified (but very general) mathematical model of the neuronal network which is defined along this section. The model, up to an abstract reformulation, and a generalization that allows any finite dimension for the impulsive differential equation governing each neuron, is taken from \cite{Izhikevich_PulseCoupledDynamicsExcitabilityBursting} and \cite{CatGuiraud}.  In the following subsections we describe   the mathematical assumptions of this model:

\vspace{-.4cm}

\subsection {Model of an isolated neuron} \label{subsectionModelEachNeuron}

\vspace{-.2cm}

{Each   neuron $i$},   while it does not receive synaptical  actions from the other cells of the network, and while its membrane potential is lower than a (maximum) \em threshold level $\theta_i>0$, \em and larger than a lower bound $L_i<0$, is assumed to be governed by a finite-dimensional differential equation of the form
\begin{equation}
\label{eqn001}\frac{dx_i}{dt} = f_i(x_i) \ \ \mbox{ if } L_i \leq x_{i,1} < \theta_i,\end{equation}
where $t$ is time, $x_i$ is a finite-dimensional vector $(x_{i,1}, \ldots, x_{i, k})$ whose components   are   real variables that describe the instantaneous state of the cell $i$, and $f_i: \mathbb{R}^k \mapsto \mathbb{R}^k$ is a Lipschitz continuous  function giving the  velocity vector $dx_i/dt$ of the changes in the state of the cell $i$, as a function of its instantaneous vectorial value $x_i(t)$. The function $f_i$ is the so called \em vector field \em  in the phase space of the cell $i$. This space is assumed to be a finite dimensional \em compact \em manifold. The advantages of considering that $\mbox{dim}(x_i) \geq 1$ (not necessarily 1) are, among others, the possibility of showing dynamical bifurcations between different rythms and oscillations that appear in some biological neurons \cite{Coombes2013},   that would not appear if the mathematical model of all the neurons were necessarily one-dimensional.

One of the components of the vectorial state variable $x_i$  (which with no loss of generality we take as the first component $x_{i,1}$) is the instantaneous membrane potential $x_{i,1}(t) = V_i(t)$ of the cell $i$.

In the sequel, we denote $x_i(t_1^-) = \lim_{t \rightarrow t_1^-} x_i(t)$ and $x_i(t_1^+) = \lim_{t \rightarrow t_1^+} x_i(t).$

In addition to the differential equation (\ref{eqn001}), it is assumed the following {\bf spiking condition} \cite{SpikingNeuronModels}:  If there exists an instant $t_1$ such that the potential $V_i (t_1^-) = x_{i,1}(t_1^-)$   equals    the \em threshold level \em $\theta_i$, then $x_{i,1}(t_1^+) = 0$. In brief, the following logic assertion holds, by hypothesis:

\vspace{-.3cm}

\begin{equation} \label{eqn002} x_{i,1}(t_1^-) = \theta_i \ \Rightarrow \  x_{i,1}(t_i^+) = 0.\end{equation}

\vspace{-.2cm}

\noindent Here, $0$ is the reset value.  It is normalized to be zero after a   change of variables, if necessary, that refers   the difference of membrane potential  of the cell $i$ to the reset value. A more realistic model   would consider a positive relatively short time-delay $\Delta t_1$ between the instant $t_1$ when the membrane potential arrives to the threshold level $\theta_i$, and the instant $t_1 + \Delta t_1$ for which the potential takes its reset value $0$. During this short time-delay, the membrane potential shows an abrupt pulse of large   amplitude, which is called \em spike \em of the neuron $i$. The impulsive simplified model approximates the spike to an abrupt discontinuity jump, by taking the time-delay $\Delta t_1$ equal to zero. Then, the spike becomes  an instantaneous jump of   the membrane potential $x_{i,1}(t)$    from the level $\theta_i \neq 0$ to the reset value $0$  which occurs  at $t= t_1$  according to condition (\ref{eqn002}).

We denote by $\delta_{\theta_i}(x_{i,1})$  the Dirac delta supported on $\theta_i$. Namely $-\int_t  \theta_i \,  d   \delta_{\theta_i(t)} (x_{i,1}) $ (via the abstract integration theory with respect to the Dirac delta probability measure) denotes a discontinuity step   $-\theta_i$ that occurs on the potential $x_{i,1}(t)$ at each instant $t = t_1$ such that   $x_{i,1}(t_1^-) = \theta_i$. In other words: $$x_{i,1}(t_1^+) - x_{i,1}(t_1^-) = -\theta_i, $$ and so $$x_{i,1}(t_1^+) = x_{i,1}(t_1^-) - \theta_i= \theta_i - \theta_i= 0.$$

After the above notation is adopted, the dynamics of each cell $i$ (while isolated from the other cells of the network) is modelled by the following \lq\lq impulsive differential equation\rq\rq:
\begin{equation}
\label{eqn003}
\frac{dx_i}{dt} = F_i(x_i), \mbox{ where } $$ $$F_i(x_i) = f_i(x_i) - \vec \theta_i \delta_{\theta_i}(x_{i,1}).
\end{equation}
In the above equality  $\vec \theta_i = (\theta_i, 0, \ldots, 0)$ is the   jump  vector with dimension equal to the dimension of the state variable $x_i$. Namely, at each spiking instant, only   the first component $x_{i,1}$ (the membrane potential) is abruptly reset, since the jump vector has all the other components equal to zero.

Strictly talking, the equation (\ref{eqn003}) is not a differential equation, but the hybrid between the differential equation $dx_i/dt = f_i(x_i)$ plus a rule, denoted by $dx_i/dt = -\vec \theta_i \delta_{\theta_i}(x_{i,1})$. This impulsive rule   imposes a discontinuity jump of amplitude vector $-\vec \theta_i$ in the dependence of the state variable $x_i(t)$ on $t$.  Therefore,   $x_i(t)$ is not  continuous, and thus it is not indeed differentiable. It is in fact discontinuous at each instant $t= t_1$ such that $x_{i,1}(t_1^-) = \theta_i$, i.e. when the Dirac delta $\Delta_{\theta_i}(x_{i,1})$ is not null.

Nevertheless, the theory of impulsive differential equations follows similar rules than the theory of ordinary differential equations. It was early initiated by Milman and Myshkis \cite{ImpulsiveDifferentialEquations}, cited in \cite{ImpulsiveDifferentialEquations2}. In particular, the existence and uniqueness of solution  for each initial condition, and   theorems of stability, still hold for the impulsive differential equation (\ref{eqn003}), as if it were an ordinary differential equation \cite{ImpulsiveDifferentialEquations,ImpulsiveDifferentialEquations2}.

\vspace{-.4cm}

\subsection {Model of the synaptical interactions among the neurons} \label{subsectionModelSynapsis}

\vspace{-.2cm}

 {The synaptical interactions} are modelled by the following rule:
 If the membrane potential $x_{i,1}$ of some neuron $i$ arrives to (or exceeds) its threshold level $\theta_i$ at instant $t_1$, then the cell $i$ sends an action $\vec \Delta_{i,j}$ to   the other neurons $j \neq i$. In particular $\vec \Delta_{i,j}$ may be zero  if no synaptical connection exists from the cell $i$ to the cell $j$. This action produces a discontinuity jump in the membrane potential  $x_{j,1}$. We   denote by $\Delta_{i,j}$   the signed amplitude of the  discontinuity  jump on the membrane potential $x_{j,1}(t)$ of the neuron $j$, which is produced by the synaptical action from the neuron $i$, when $i$ spikes. The real value $\Delta_{i,j}$ may depend on the instantaneous state $x_j$ of the receiving neuron $j$ just before the synaptic action from neuron $i$ arrives. For simplicity we do not explicitly write this dependence. Thus, the symbol $\Delta_{i,j}$ denotes a real function of $x_j$, which we assume to be   either identically null or with constant sign.

  We denote by $\vec \Delta_{i,j} = (\Delta_{i,j}, 0, \ldots, 0)$   the   discontinuity jump vector, with dimension equal to the dimension of the variable state $x_j$ of the cell $j$. In other words, the discontinuity jump in the  instantaneous vector state $x_j$ of the cell $j$,  that is  produced when the cell $i$ spikes, is null on all the components of $x_j$ except the first one $x_{j,1}$, i.e. except on the membrane potential of the neuron $j$.   In formulae:
\begin{equation}
\label{eqn004}   x_{i,1}(t_1^-) = \theta_i \ \Rightarrow \ x_{j,1} ( t_1^+) = x_{j,1}(t_1^-) + \Delta_{i,j}$$ $$ \ \  x_j (t_1^+) = x_j(t_1^-) + \vec \Delta_{i,j}.
\end{equation}
 Thus, the dynamics of the whole neuronal network is modelled by the following system of impulsive differential equations:
\begin{eqnarray}
\frac{dx_1}{dt} & = & F_1(x_1)  + \sum_{i \neq 1} \vec\Delta_{i,1} \delta_{\theta_i}(x_{i,1}), \nonumber\\
\frac{dx_2}{dt} & = & F_2(x_1)   + \sum_{i \neq 2} \vec\Delta_{i,2} \delta_{\theta_i}(x_{i,1}), \nonumber\\
\ldots &   &  \nonumber\\
\frac{dx_j}{dt} & = & F_j(x_j)   + \sum_{i \neq j} \vec\Delta_{i,j} \delta_{\theta_i}(x_{i,1}), \label{eqn005}\\
\ldots &   &  \nonumber \\
\frac{dx_N}{dt} & = & F_N(x_N)   + \sum_{i \neq N} \vec\Delta_{i,N} \delta_{\theta_i}(x_{i,1}), \nonumber
\end{eqnarray}
where $N$ is the number of cells in the network.

\vspace{-.2cm}

\begin{definition}
\label{definitionExcitInhibitMixed} {\bf (Excitatory, inhibitory and mixed neurons)} \em
The synapses from cell $i$ to $j$ is called excitatory if $\Delta_{i,j}>0$ and it is called inhibitory if $\Delta_{i,j} < 0$. If $\Delta_{i,j}= 0$ then there does not exist synaptical action from the cell $i$ to the cell $j$. A neuron $i$ is called \em excitatory \em (resp. \em inhibitory\em) if $\Delta_{i,j} >0$ (resp. $\Delta_{i,j} < 0$) for all $j$ such that $\Delta_{i,j} \neq 0$. The cell $i$ is called \em mixed \em if it is neither excitatory nor inhibitory. Dale's Principle (which we do not assume a priori to hold) states that no neuron is mixed. \end{definition}

%\vspace{-.8cm}

\begin{remark}
 \label{remarkNoIndifferentCells} \em
 It is not restrictive to   assume that no cell  $i$ is \em indifferent\em, namely no cell $i$ sends null synaptical actions to all the other cells, i.e.

 \vspace{-.3cm}

  $$\not \! \exists \ i \in \{1, 2, \ldots, m\} \mbox{ such that } \Delta_{i,j} = 0 \ \forall j \neq i.$$

 \vspace{-.3cm}

  \noindent In fact, if there existed such a cell $i$, it would not send any action to the other cells of the network ${\mathcal N}$. So, the global dynamics of the network is not modified (except for having one less variable) if we take out the cell $i$ from ${\mathcal N}$.

  \noindent \em All along the paper we assume that the network ${\mathcal N}$ has at least 2 neurons and no   neuron is indifferent. \em \end{remark}

 \vspace{-.8cm}

\subsection {The refractory rule} \label{subsectionModelRefractAvalanche}

\vspace{-.2cm}

\noindent To obtain a well defined   deterministic dynamics from the system (\ref{eqn005}), other complementary assumptions are adopted by the model. First, a {\bf refractory phenomenon } (see for instance
 \cite[page 725]{refractory})
 is considered as follows: If some fixed neuron $j$ spikes at instant $t_1$, then its potential $x_{j,1}$ is reset to zero becoming indifferent to the synaptical actions that it may receive  (at the same instant $t_1$) from the other neurons. Second, if  for some fixed neuron $j$ at some instant $t_1$, the sum $\sum_{i \neq j} \max\{0, \Delta_{i,j}\}$ of the excitatory actions that $j$ simultaneously receives from the other neurons of the network, is larger or equal than $\theta_j - x_{j,1}(t_1^-)$, then $j$ itself spikes at instant $t_1$, regardless whether $x_{j,1}(t_1^-) = \theta_i$ or not. In this case, at instant $t_1$ the cell $j$ sends synaptical actions $\Delta_{j,h}$ to the other neurons $h$ of the network, and then, the respective potentials $x_{h,1}$ will suffer a jump $\Delta_{j,h}$ at instant $t_1$. This process may make new neurons $h$ to spike  in an {\bf avalanche process} (see \cite{CatGuiraud}). This avalanche   is produced instantaneously, when some excitatory neuron spontaneously arrived to its threshold level. But due to the refractory rule, once each neuron   spikes, its membrane potential refracts all the excitations or inhibitions that come at the same instant. So, the avalanche phenomenon  is produced instantaneously,  \em but includes each neuron $i$ at most once. \em  Then, each interaction term $\vec \Delta_{i,j} \delta_{\theta_i}(x_{i,1})$ in the sum at right of Equation (\ref{eqn005}) is added  only once   at each spiking instant $t_1$.

\vspace{-.6cm}

\section{\bf   First step  of the proof \\ (Graphs, parts and units)}

\label{sectionGraph&Layers}

\vspace{-.3cm}

The purpose of this section is to prove Proposition \ref{propositionUnits} and to state the existence of an \lq\lq Inter-units Graph\rq\rq \ (Definition \ref{definitionInter-UnitsGraph}). These are   intermediate results (the first step) of the proof of Main Theorems \ref{theoremDalePrinciple} and \ref{theoremCounterexample}. We will prove these intermediate results  by logical deduction  from several previous statements and hypothesis. So, we start by  including the needed previous statements in the following series of mathematical definitions:

Let ${\mathcal N}$ be a network of $N \geq 2$ neurons, according to the model defined in Section \ref{sectionModel}.
\begin{definition}
\label{definitionGraph} {\bf (The network's graph)} \em
We call a directed and weighted graph $G$ \em the graph of the network ${\mathcal N}$ \em if the  vertices  of $G$ are the cells $i \in \{1, 2, \ldots, N\}$ of ${\mathcal N}$,   each   edge  $e_{i,j}, \   i \neq j$ of $G$, corresponds to each   nonzero synaptical action  from the cell $i$ to the cell $j$ and conversely, and $e_{i,j}$ has   weight   $\Delta_{i,j} \neq 0$.  (See the example of Figure \ref{Figure1}.)

\noindent To unify the notation, we agree:

\noindent $\bullet$ ${\mathcal N}$   denotes either the network or its graph;

 \noindent  $\bullet$ $i$ is either a cell of ${\mathcal N}$ or a node of the graph;

  \noindent  $\bullet$  $\Delta_{i,j}$ denotes either the synaptical action from $i$ to $j$,  or the weight of the edge $e_{i,j}$ in the graph, or this edge itself.
\end{definition}

\begin{definition}
\label{definitionIdenticalCells} {\bf (Structurally identical cells)} \em Two different cells $i \neq j$ are \em structurally identical \em if $F_i = F_j$ in the respective differential equations (\ref{eqn003}), $\Delta_{i,j}= \Delta_{j,i} = 0$, and $\Delta_{h,i} = \Delta_{h,j}$ for all $h \neq i,j$.   These conditions imply that the dynamical systems that governs neurons $i$ and $j$  are the same. So, their future dynamics may differ \em only \em because their initial states \em $x_{i}(0)$ and $x_j(0)$ may be different. \em Note that, if $i$ and $j$ are structurally identical, then by definition,    the edges of the graph at the receiving  nodes $i$ and $j$ (from any other fixed sending node $h$) are respectively equally weighted by $\Delta_{h,i}= \Delta_{h,j}$. Nevertheless,  the edges from $i $ and $j$, as sending nodes of the network, are not necessarily identically weighted, i.e. $\Delta_{i,h}$ may be different from $\Delta_{j,h}$.

 In Figure \ref{Figure1} we represent a graph $G$ with three mutually identical cells $1, 2 $ and $3$, provided that $F_1 = F_2 = F_3$ in the equation (\ref{eqn003}) and $\Delta_{h,1} = \Delta_{h,2}= \Delta_{h,3}$ for $h= 4, 5$. Besides, the graph $G$ has two other nodes, which corresponds to the neurons $4$ and $5$. The cells $4$  and $5$ are not mutually identical  because the synaptical actions that they receive from  the other cells are not equal.
\end{definition}

{\begin{figure}
[h]
\begin{center}
\vspace{-.7cm}
\includegraphics[scale=.4]{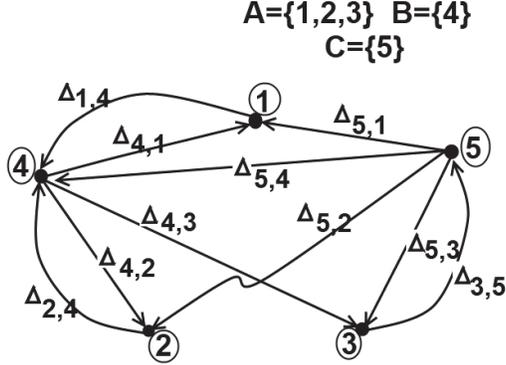}
\vspace{0cm}
\caption{\label{Figure1} The graph of a network ${\mathcal N} = \{1,2,3,4,5\}$. The directed and weighted edges correspond to the nonzero synaptical interactions $\Delta_{j,i}$ among the neurons $j \neq i$.}
\vspace{-.5cm}
\end{center}
\end{figure}}

The above definitions and the following ones are just \em mathematical tools,  \em with no other purpose than enabling us to prove Theorems \ref{theoremDalePrinciple} and \ref{theoremCounterexample}. They are not aimed to explain   physiological or functional roles of subsets of real biological neurons in the brain or in the nervous system. Nevertheless, it is rather surprizing that the following abstract mathematical tools, which we  include here just to prove Theorems  \ref{theoremDalePrinciple} and \ref{theoremCounterexample}, have indeed a  resemblance with   concepts or phenomena that are   studied by Neuroscience. In particular, the following Definitions \ref{definitionHomogeneousPart} and \ref{definitionSynapticalUnit} of homogeneous part and synaptical unit of a neuronal network, are roughly analogous to  the concepts of regions, subnetworks or groups of many similar neurons,  characterized by  a certain structure and  a collective physiological role. For instance some subnetworks or layers of biological or artificial neurons are defined according to the role of their synaptical interactions with other subnetworks or layers \cite{DistributionInhibitoryExcitatotySynapsesOnHippocampalCells}.

\begin{definition} \label{definitionHomogeneousPart}
{\bf (Homogeneous Part)} \em An \em homogeneous part  \em of the neuronal  network is a maximal subset of cells of the network that are mutually pairwise identical (cf. Definition \ref{definitionIdenticalCells}). As a particular case, we agree to say that an homogeneous part is composed by a single neuron $i$ when no other neuron is structurally identical to $i$. In Figure \ref{Figure1} we draw the graph of a network composed by three homogeneous parts $A, B$ and $C$. The homogeneous part $A$ is composed by the three identical neurons 1, 2 and 3, provided that $F_1 = F_2 = F_3$ and $\Delta_{h,1}=   \Delta_{h,2} = \Delta_{h,3}$ for $h= 4, 5$. The homogeneous parts $B = \{4\}$ and $C= \{5\}$ have a single neuron each because $\Delta_{h, 4} \neq \Delta_{h,5}$ for some $h$ (for instance for $h= 2$).
\end{definition}

\begin{definition}
\label{definitionSynapticalUnit}{\bf Synaptical Unit} \em A  \em synaptical unit \em is a subset $U \subset A$ of an homogeneous part $A$ of a neuronal network such that:

\noindent $\bullet$ For any neuron $h \not \in A$   there exists \em at most \em one neuron $i \in U$ such that $\Delta_{i,h} \neq 0$.

\noindent $\bullet$ $ A$ is partitioned in a minimal number of sets $U$ possessing the above property.

In particular, a synaptical unit may be composed by a single neuron. This occurs, for instance, when for some neuron $h \not \in A$ and for any neuron $i \in A$ the synaptical interaction $\Delta_{i,h}$ from $i$ to $h$ is nonzero.  In Figure \ref{Figure1} we draw the graph of a network composed by three homogeneous parts $A$, $B$ and $C$ such that: $A$ is composed by three   identical neurons 1, 2 and 3, that form two synaptical units $U_1 := \{1\}$ and $U_2 = U_3:= \{2,3\}$. In fact, the cells $1$ and $2$ can not belong to the same unit because there exist nonzero actions departing from both of them to neuron $4$.  One can also form the two synaptical units of $A$ by defining $U_1 = U_3:= \{1,3\}$ and $U_2:= \{2\}$.   The homogeneous part  $B$ is composed by a single neuron $4$, and thus, it is    a singe synaptical unit $U_4 = \{4\} = B$. Analogously $C$ is composed by a single neuron $5$, and thus it is a single synaptical unit $U_5 := \{5\} = C$. The total number of neurons of the network in Figure \ref{Figure1} is 5, the total number of synaptical units is 4,    the total number of homogeneous parts is 3,   the total number of nonzero synaptical interactions among the neurons is 9, but the total number of synaptical interactions among different homogeneous parts is only 5 (see Figure \ref{Figure2}).

When  a synaptical unit $U$ has more neurons,     the following quotient $Q_U $ diminishes:  $Q_U$ is the   number of synaptical connections departing from the cells of $U$  divided by the total number of neurons of $U$. In fact, by Definition \ref{definitionSynapticalUnit},  for each synaptical unit $U$ there exists \em at most \em one nonzero synaptical action to any other fixed  neuron  $h$ of the network, regardless how many cells compose $U$. So, if we enlarge the number of cells in $U$,   the number of nonzero synaptical  actions departing from the cells of $U$ remains constant.  Thus, the quotient $Q_U$ diminishes. Although this quotient $Q_U$ becomes smaller when  the number of neurons of the synaptical unit $U$ enlarges,   in     Theorem \ref{theoremExistenceofOptimalNetworks}   we will rigourously prove the following result:

The dynamical system governing a neuronal network ${\mathcal N}$ with the maximum number of neurons in each of its synaptical units, \em is the richest one, \em  i.e. ${\mathcal N}$ will exhibit the largest  set of different orbits in the future, and so it will be theoretically capable to perform the most diverse set of processes.
\end{definition}

\vspace{-.3cm}

The following result proves that any neuronal network, according to the mathematical model of Section \ref{sectionModel}, is decomposed as the union of at least two homogeneous parts, and each of these parts is decomposed into pairwise disjoint synaptical units. It also states  the existence of an upper bound for the number of   neurons  that  any    synaptical unit can have.

\vspace{-.3cm}

\begin{proposition} {\bf (Intermediate result in the proof of Main  Theorems  \ref{theoremDalePrinciple} and \ref{theoremCounterexample})}

\label{propositionUnits} Let ${\mathcal N}$ be any network according to the mathematical model defined in Section \em \ref{sectionModel}. \em Then:

 \noindent {\bf (i) } The set of neurons of ${\mathcal N}$ is the union of exactly $l \geq 2$ pairwise disjoint homogeneous parts.

  \noindent {\bf (ii) } Each homogeneous part $A$  is the union of a positive finite number of pairwise disjoint synaptical units.

   \noindent {\bf (iii) } The total number  of neurons of each synaptical unit   is   at least one and at most    $l-1$.

   \noindent {\bf (iv)} For each synaptical unit $U$ and for each homogeneous part $B$  there exists a unique real number $\Delta_{U,B}$ that satisfies  the following properties:

     $\Delta_{U, B}= 0$ if and only if  $\Delta_{i,h} = 0$ for all $i \in U$ and for all $h \in B$. In particular $\Delta_{U, A}= 0$ if $U \subset A$.

    $\Delta_{U, B} \neq 0$ if and only if $\Delta_{i,h} = \Delta_{U, B}$ for one and only one neuron $i \in U$ and for all $h \in B$, and $\Delta_{j, h} = 0$ for all $h \in B$ and all $j \in U$ such that $j \neq i$.
\end{proposition}
{\em Proof: } (i) We denote $i \equiv j$ if the cells $i$ and $j $ are structurally identical according to   Definition \ref{definitionIdenticalCells}. We add the rule  $i \equiv i$ for any cell $i$. Thus, $\equiv$ is an equivalence relation. From Definition \ref{definitionHomogeneousPart} the $\equiv$ classes of neurons are the homogeneous parts of the network.

Since the equivalence classes of any equivalence relation in any set determine a partition of this set, then the network, as a set of neurons, is the union of its pairwise disjoint homogeneous parts. Denote by $l \geq 1$ the total number of different homogeneous parts that compose the network. Let us prove that $l \geq 2$. In fact, if $l$ were equal to 1, then, by Definition \ref{definitionIdenticalCells}, $\Delta_{i,j} = 0$ for any pair of cells, contradicting the assumption that no cell is indifferent (see the end of Remark \ref{remarkNoIndifferentCells}). We have proved Assertion (i).

\noindent (ii) Fix an homogeneous part $A$, and fix some neuron $i  \in A$. Consider the set of neurons $S_{i} := \{ h: \ \Delta_{i, h} \neq 0\}$. The set $S_{i}$ is nonempty because the cell $i $ is not indifferent (see   Remark \ref{remarkNoIndifferentCells}). Choose and fix a neuron $h \in S_{i}$. We discuss two cases: either $h \in S_{j}$ for all $j \in A$, or the set $\{j \in A: h \not \in S_j\}$ is nonempty.

In the first case, for each neuron $j \in A$ the singleton $\{j\}$ (formed by the single element $j$), satisfies Definition \ref{definitionSynapticalUnit}. Thus, $\{j\}$ is a synaptical unit
for all $j \in A$ and assertion (ii) is  proved.

In the second case, consider the set  $A':= \{i\} \bigcup \{j \in A: h \not \in S_j\} \supset \neq \{i\}$. Consider also (if they exist)  all the singletons $ \{j\}$ where $j \in A $ is such that $h \in S_j$. These latter sets $\{j\}$ satisfy Definition \ref{definitionSynapticalUnit} and, thus, they are pairwise disjoint synaptical units, which are also disjoint with $A'$. Besides, their union with $A'$ compose     $A$. So, it is now enough to prove that $A'$ is also the union of pairwise disjoint synaptical units.

 Now, we choose and fix a neuron $i' \in A'$ such that $i' \neq i$. (Such a neuron exists because $A' \neq \{i\}$). By construction of the set $A'$, we have $h \not \in S_{i'}$. But, since the neuron $i' $ is not indifferent, there exists $h' \in S_{i'}$. So, we can repeat the above argument putting $i'$ in the role of $i$, $h'$ in the role of $h$, and $A'$ in the role of $A$.

 Since the number of neurons is finite, after a finite number of steps  (repeating the above argument at each step), we obtain a decomposition of $A$ into a finite number of pairwise disjoint sets that are synaptical units, ending the proof of Assertion (ii).

\noindent (iii) Let $U$ be a synaptical unit. By Definition \ref{definitionSynapticalUnit}, $U \subset A$ where $A$ is an homogeneous part of the set of neurons. By Assertion (i) there are   exactly $l-1 \geq 1$ other homogeneous parts  $B \neq A$.   From Definitions \ref{definitionIdenticalCells} and \ref{definitionHomogeneousPart}, for any fixed $i \in U$:  $\Delta_{i,h} = \Delta_{i, h'} $ for all $h,h' \in B$. So, for each  $B$ we denote $$\Delta_{i, B} := \Delta_{i,h} \ \forall \ h= \in B.$$  Since any neuron $i \in U \subset A  $ is not indifferent,  there exists at least one homogeneous part $B \neq A$ such that $\Delta_{i, B} \neq 0$. Besides, applying Definition \ref{definitionSynapticalUnit}, for each homogeneous part $B \neq A$, there exists at most one neuron $i \in U$ such that $\Delta_{i, B} \neq 0$. The last two assertions imply that there is a one-to-one correspondence (which is not necessarily surjective) from the set of neurons in $U$ to the set of homogeneous parts that are different from $A$. Then, the number of neurons in $U$ is not larger than the number of existing homogeneous parts $B \neq A$, i.e. it is not larger than $l-1$. We have proved Assertion (iii).

\noindent (iv) Fix an arbitrary synaptical unit $U \subset A$ (where $A$ is the homogeneous part that contains $U$) and an arbitrary homogeneous part $B$ (in particular, $B$ may be $A$). As in the  above proof of Assertion (iii), for each neuron $i \in U$ it is defined $\Delta_{i, B}$ such that $\Delta_{i, h} = \Delta_{i, B}$ for all $h \in B$. By Definition \ref{definitionSynapticalUnit}, either $\Delta_{i, B} = 0$ for all $i \in U$, or $\Delta_{i, B} \neq 0$ for one and only one cell $i \in U$. In the first case we define $\Delta_{U, B}= 0$ and in the second case we define $\Delta_{U, B} = \Delta_{i, B}$. By construction, Assertion (iv) holds: in particular, from Definitions \ref{definitionIdenticalCells} and \ref{definitionHomogeneousPart}, we have $\Delta_{i,h} = 0$ for all $i \in U \subset A$ and for all $h \in A$. So, $\Delta_{U, A}= 0$ if $U \subset A$.
\hfill $\Box$

\begin{definition}
{\bf (Inter-units graph - Intermediate result in the proof of Theorems \ref{theoremDalePrinciple} and \ref{theoremCounterexample})} \label{definitionInter-UnitsGraph}
\em As a consequence of Proposition \ref{propositionUnits} the graph of a neuronal network ${\mathcal N}$ can be represented by a  simpler   one, which we call \em the inter-units graph\em. This is, by definition, the graph whose nodes are not the cells but the synaptical units. Each directed and weighted edge  in the inter-units graph, links a synaptical unit  $U \subset A$ with the synaptical unit $V \subset B$ ($A \neq B$). It is weighted by the synaptical action  $\Delta_{U, B} \neq 0$. For instance, the network ${\mathcal N}$ of Figure \ref{Figure1} is   represented by the inter-units graph of Figure \ref{Figure2}.
\end{definition}
{\begin{figure}
[h]
\begin{center}
\vspace{-.5cm}
\includegraphics[scale=.5]{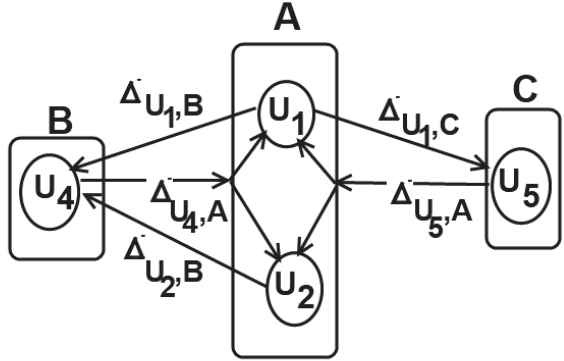}
\vspace{-.5cm}
\caption{\label{Figure2} The inter-units graph of the network ${\mathcal N}  = \{1,2,3,4,5\}$ of Figure \ref{Figure1}. It is composed by three homogeneous parts $A$, $B$ and $C$. The part $A$ is composed by two synaptical units $U_1 = \{1,3\}$ and $ U_2 = \{2\}$, the part $B$ is the single unit   $U_4 = \{4\}$ and the part $C$ is the single unit $U_5 = \{5\}$.}
\vspace{-.6cm}
\end{center}
\end{figure}}

\noindent{\bf Interpretation:} The inter-units graph of a neuronal network, according to Definition \ref{definitionInter-UnitsGraph}, recovers the \em essential \lq\lq anatomy\rq\rq \em \ of the spatial distribution of the synaptical connections of the network,  among \em  groups \em of mutually identical cells    (the  so called synaptical units). This description, by means of the inter-units graph, recalls
    experimental studies on the synaptical activity of some neuronal subnetworks of the brain. For instance, in \cite{DistributionInhibitoryExcitatotySynapsesOnHippocampalCells}, Meg\'{\i}as et al. study the spatial distribution of inhibitory and excitatory synapses inside the hippocampus.

%\vspace{-.1cm}

Each synaptical unit acts, in the inter-units graph, as if it were a single neuron. The spatial statical structure of groups of   synaptical connections is the only observed object by this graph. Besides, the inter-units graph does not change if the number of neurons composing each of the many synaptical units, change. In the following section, we will condition the study of the networks  to those that have mutually  \em isomorphic  inter-units graphs, \em  i.e. they have the same statical structure of   synaptical connections among groups of identical cells.

%\vspace{-.1cm}

In Section \ref{sectionDynamicalOptimum}, we will look on the \em dynamical responses \em of the network  that have the same (statical) inter-units graph of synaptical connections.   Any change in the number of neurons will change the space of possible initial states, and so the space of possible orbits and the global dynamics. So, among all the networks that have isomorphic inter-units graphs, the network with more neurons should, a priori, exhibit a larger diversity of theoretic possible dynamical responses to external stimulus.

%\vspace{-.1cm}

For instance, two identical neurons $1$ and $2$ in a synaptical unit $U$ define a space of initial states (and so of orbits) that is composed by all the pairs $(x_1(0), x_2(0))$ of vectors in the phase space of each neuron. But three identical neurons $1$, $2$ and $3$ in $U$, define a space of initial states composed by all the triples $(x_1(0), x_2(0), x_3(0))$ of vectors. So,   the diversity of orbits that a neuronal network   can exhibit, enlarges if  the
number of neurons of each synaptical unit enlarges.  In Section \ref{sectionDynamicalOptimum}, we will study the theoretical optimum in the dynamical response of a family of networks that are synaptical equivalent. We will prove that this optimum exists and that it is   achieved when the network has the maximum number of cells  (Theorem \ref{theoremExistenceofOptimalNetworks}).

\vspace{-.5cm}

\section{Second step of the proof \\(Synaptical equivalence between networks)} \label{sectionEquivalentNetworks}

\vspace{-.3cm}

The purpose of this section is to prove the existence of an equivalence relation (Definition \ref{definitionEquivalentNetworks}) in the space of \em   all \em the neuronal networks   modelled by the mathematical hypothesis of Section \ref{sectionModel}. This is  the   intermediate result  in the second step of the proof of Main Theorems \ref{theoremDalePrinciple} and \ref{theoremCounterexample}. We will deduce this intermediate  result from the previous ones obtained in Section \ref{sectionGraph&Layers}.

Let ${\mathcal N}$ and ${\mathcal N'}$ be two neuronal networks according to the model defined in Section \ref{sectionModel}. Denote:

\noindent $N$ and $N'$ the numbers of neurons of ${\mathcal N}$ and ${\mathcal N}'$ respectively,

\noindent $i$ and $i'$ a (general) neuron of ${\mathcal N}$ and ${\mathcal N}'$ respectively,

\noindent $l$ and $l'$ the respective numbers of homogeneous parts of ${\mathcal N}$ and ${\mathcal N}'$, according to Definition \ref{definitionHomogeneousPart}.

\noindent  $s$ and $s'$ the  respective numbers of synaptical units according to Definition \ref{definitionSynapticalUnit}.

\noindent $B$ and $B'$   a (general) homogeneous part of ${\mathcal N}$  and   ${\mathcal N}'$  respectively.

\noindent $U$ and $U'$   a (general) synaptical unit of ${\mathcal N}$  and   ${\mathcal N}'$  respectively.

 \noindent $\Delta_{U, B}, \ \ \Delta_{U', B'}$  the synaptical weights, according to part (iv) of Proposition \ref{propositionUnits},  of ${\mathcal N}$  and   ${\mathcal N}'$  respectively.

\begin{definition}
{\bf (Synaptically equivalent networks - Intermediate result in the proof of Main Theorems \ref{theoremDalePrinciple} and \ref{theoremCounterexample})} \em \label{definitionEquivalentNetworks}

\noindent We say that ${\mathcal N}$  and ${\mathcal N'}$ are \em synaptically equivalent \em if:

\noindent $\bullet$ $l= l', \ s= s'$, according to the above notation.

\noindent $\bullet $ There exists a one-to-one and surjective correspondence $\varphi$ from the set  of synaptical units $U$ of ${\mathcal N}$ and the set of synaptical units $U' = \varphi (U)$ of ${\mathcal N}'$  such that
$$\Delta_{U, B} = \Delta_{U', B'} \ \ \forall \ (U, B), $$
where   $B' = \varphi(B)$ is   the homogeneous part of the network ${\mathcal N}'$ whose synaptical units are the images by $\varphi$ of the synaptical units that compose $B$.

\noindent $\bullet$ For any synaptical unit $U$ of ${\mathcal N}$
$$F_i = F_{i'} \ \ \forall \ i \in U, \ \ \forall \ i' \in U'= \varphi(U),$$
where $F_i$ and $F_{i'}$ are the second terms of the impulsive differential equations (\ref{eqn003}) that govern the dynamics of the neurons $i$ and $i'$, respectively.

In other words, the networks ${\mathcal N}$ and ${\mathcal N}'$ are synaptically equivalent, if there exists an isomorphism $\varphi$ between their respective inter-units graphs such that all the neurons in the unit $U$ of the network ${\mathcal N}$ are structurally identical to all the neurons in the unit $\varphi(U)$ of the network ${\mathcal N}'$.
\end{definition}

\begin{figure}
[h]
\begin{center}
\vspace{-.4cm}
\includegraphics[scale=.4]{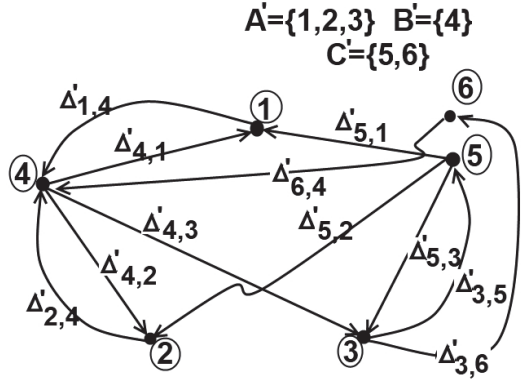}
\end{center}
\vspace{0cm}
\caption{\label{Figure3} {The graph of a network ${\mathcal N}' = \{1,2,3,4,5,6\}$. It is composed by three homogeneous parts $A'$, $B'$ and $C'$. This network ${\mathcal N}'$} is synap\-tic\-ally equivalent to the network ${\mathcal N}$ of Figure \ref{Figure1}.}

\vspace{-.3cm}

\end{figure}

For example, let us consider the network ${\mathcal N}'$ of Figure \ref{Figure3}. Assume that $F_1 = F_2 = F_3$, \ $F_5 = F_6$ in the respective impulsive differential equations (\ref{eqn003}). Also assume that $\Delta'_{h,1}= \Delta'_{h,2} = \Delta'_{h,3}   $ for $h= 4,5$ and $\Delta'_{3,5} = \Delta'_{3,6}$. Then, the network ${\mathcal N}'$ has $3$ homogeneous parts $A'= \{1,2,3\}, \ B'= \{4\}, \ C'= \{5,6\}$. Analogously to the example of Figure \ref{Figure1}, the part $A'$ of the network ${\mathcal N}'$ is composed by two synaptical units $U'_1 = U'_3 = \{1,3\}$ and $U'_2 = \{2\}$, and the part $B'$ is composed by a single synaptical unit $U'_4 = \{4\}$. Finally, the part $C'$ of the network of Figure \ref{Figure3} is composed by a single synaptical unit $U'_5 = U'_6 = \{5,6\} $.

Assume that the interactions $\Delta_{i,j}$  and $\Delta_{i,j}'$ of the networks ${\mathcal N}$ and ${\mathcal N}'$ of Figures \ref{Figure1} and \ref{Figure3} respectively, satisfy the following properties:

\noindent $\Delta_{i,j} = \Delta'_{i , j }$ for all $i, j \in\{1,2,3,4\}$ such that $i \neq j$,

\noindent $\Delta_{5,j} = \Delta'_{5,j}$ for all $j\in \{ 1,2,3\}, $

\noindent $\Delta_{3, 5} = \Delta'_{3,5} = \Delta'_{3,6}$,

\noindent $\Delta_{5,4} = \Delta'_{6,4}$.

Then, the inter-units graph of Figure \ref{Figure2} also corresponds to the network ${\mathcal N}'$. So, the networks ${\mathcal N}$ and ${\mathcal N}'$ of Figures \ref{Figure1} and \ref{Figure3} are synaptically equivalent.

We note that, for synaptically equivalent networks, the number of neurons, and   also the number of nonzero synaptical interactions, may vary. For instance, the networks of Figures \ref{Figure1} and \ref{Figure3} are synaptically equivalent, but their respective total numbers of of neurons and  of synaptical interactions are mutually different.

\noindent {\bf Comments: } The equivalence relation between   networks ${\mathcal N}$ and ${\mathcal N}'$, according to Definition \ref{definitionEquivalentNetworks},   implies that both ${\mathcal N}$ and ${\mathcal N}'$ will have exactly the same dynamical response (i.e. they will follow the same orbit),  \em  provided that,  for any dynamical unit $U$ of ${\mathcal N}$,   the initial states of  {\bf \em all } the neurons  in $U$  are mutually equal and also equal to the initial states of  {\bf  \em all} the neurons   in the dynamical unit $U' = \varphi (U)$   of the other network. \em In fact, since the impulsive differential equations (\ref{eqn003}) that govern  the dynamics of all those neurons coincide, and since the synaptical jumps that each of those neurons receive from the other neurons of its respective network    also coincide, their respective deterministic orbits in the future must coincide if the initial states are all the same.

Nevertheless, if not all those initial states are mutually equal, for instance if some external signal changes the instantaneous states of some but not all the neurons in a synaptical unit, then their respective orbits will differ, during at least some finite interval of time. In this sense, each synaptical unit with more than one neuron, is a group of identical cells that \em distributes \em the dynamical process among its cells, i.e. \em it has the capability of dynamically distributing the information. \em

In brief, two synaptically equivalent networks have, as a common feature, the same statical configuration or \lq\lq anatomy\rq\rq \ of the synaptical interactions between their units (i.e. between groups of identical cells, equally synaptically connected). Then, both networks would evolve equally, under the hypothetical assumption that all the initial states of the neurons of their respective synaptical units coincided. But the two networks may exhibit qualitatively \em   different dynamical responses \em to external perturbations or signals, if these signals make different the instantaneous states of different neurons in some synaptical unit. Such a difference produces a diverse  distribution  of the dynamical response among the cells.

 \vspace{-.4cm}

\section{Third step of the proof \\ (Dynamically optimal networks)}  \label{sectionDynamicalOptimum}

 \vspace{-.3cm}

 The purpose of this section is to prove Proposition \ref{propositionNumberOfNeuronsLarger} and Theorem \ref{theoremExistenceofOptimalNetworks}.   These are   intermediate results (the third step) of the proof of Main Theorems \ref{theoremDalePrinciple} and \ref{theoremCounterexample}. We will prove these intermediate results by logical deduction  from several previous statements and hypothesis. So, we start by  including the needed previous statements in the following series of mathematical definitions, remarks and notation agreements:

We   condition the study to the networks of any fixed single class, which we denote by ${\mathcal C}$, of synaptically equivalent networks  according to Definition \ref{definitionEquivalentNetworks}. In this section we search for networks exhibiting an optimum dynamics conditioned to ${\mathcal C}$.

\noindent{\bf Notation: }

  We consider the mathematical model of a general neuronal network ${\mathcal N} \in {\mathcal C}$, given by the system (\ref{eqn005}) of impulsive differential equations. We denote by

 \vspace{-.4cm}

 $$X(t)= (x_{1}(t), x_{2}(t), \ldots, x_{N}(t))$$

 \vspace{-.2cm}

 \noindent the instantaneous state  of the network ${\mathcal N}$ at instant $t$, where $N \geq 2$ is the number of neurons of the network, and $x_i \in M_i$ is the   instantaneous state of the neuron $i \in \{1, 2, \ldots, N\}$. Since by hypothesis $x_i$ evolves on a finite-dimensional compact manifold $M_i$, the state $X$ of the network evolves on the finite-dimensional compact product manifold $M$, defined by the following equality:

 \vspace{-.4cm}

 $$M := M_1 \times M_2 \times \ldots \times M_N.$$

 \vspace{-.2cm}

 \noindent In other words, $M$ is the cartessian product of the manifolds $\{M_i\}_{1 \leq i \leq N}$. Then

 \vspace{-.4cm}

\begin{equation}
  \label{eqn12} \mbox{dim}(M) = \sum_{i= 1}^N \mbox{dim}(M_i)
 \end{equation}

 \vspace{-.6cm}

\begin{definition}
 \label{definitionDynamicsofN} {\bf (Dynamics of the network)} \em
Consider any initial state

 \vspace{-.4cm}

 $$X_0 = X(0) = (x_1(0), x_2(0), \ldots, x_N(0)) \in M$$

 \vspace{-.2cm}

 \noindent of the network ${\mathcal N}$, i.e. $x_i(0) \in M_i$ is the   state of the neuron $i$ at the instant $t= 0$.

   The solution $X(t), \ \ t \geq 0 $ of the system (\ref{eqn005}) of impulsive differential equations that govern the dynamics of ${\mathcal N}$,  exists and is unique,  provided that the initial condition $X(0)= X_0 \in M$ is given (see for instance \cite{ImpulsiveDifferentialEquations}, cited in \cite{ImpulsiveDifferentialEquations2}).  We denote:

 \vspace{-.4cm}

$$\Phi(X_0, t):= X(t) \ \ \mbox{ such that } \ X(0)= X_0,$$

 \vspace{-.2cm}

\noindent and call $\Phi$ the (deterministic) \em dynamical system  \em (or, in brief,  the \em dynamics\em)  associated to the network ${\mathcal N}$. It is an autonomous deterministic dynamical system.

For any autonomous deterministic dynamical system (also if it were not modelled by differential equations), we have the following properties:

 \vspace{-.4cm}

$$\Phi(X_0, 0)= X_0, $$ $$\Phi(\Phi(X_0, t), s) = \Phi(X_0, t+s) $$

 \vspace{-.3cm}

  $$ \forall \ X_0 \in M, \ \forall \ t,s \geq 0.$$

 \vspace{-.2cm}

\noindent So, for any fixed instant $T > 0$ the state $\Phi(X_0, T)$ plays the role of  a new \lq\lq initial\rq\rq \ state, from which the orbit $\{\Phi(\Phi(X_0, T), s)\}_{s \geq 0}$   evolves for time $ s \geq 0$. This orbit coincides with the piece of orbit $\{\Phi(X_0, T+ s)\}_{T+s \geq T}$ (for time $\geq T$) that had the initial state $X_0$.
\end{definition}

\vspace{-.4cm}

\begin{definition}
{\bf (Partial Order in ${\mathcal C}$)}  \label{definitionRicherNetwork}
\em

Let  ${\mathcal N}$ and ${ \mathcal N '}$ be two networks in ${\mathcal C}$ and denote by $\Phi$ and $\Phi'$ the dynamics of ${\mathcal N}$ and ${\mathcal N}'$ respectively. Denote by $M$ and $M'$ the compact manifolds where   $\Phi$ and $\Phi'$ respectively evolve.

We say that \em ${\mathcal N}$ is dynamically richer   than ${\mathcal N}'$, \em and write $${\mathcal N}'  \sqsubseteq {\mathcal N},$$ if there exists a continuous and one-to-one (non necessarily surjective) mapping $$\psi: M' \mapsto M $$
such that
\begin{equation}
\label{eqn006}
\psi (\Phi'(X'_0, t)) = \Phi(\psi (X'_0), t) \ \ \forall \ t \geq 0, \end{equation}
for any initial state $X'_0 \in M'$.

 In other words, ${\mathcal N}' \sqsubseteq {\mathcal N}$ if and only if the dynamical system $\Phi'$ of ${\mathcal N}'$ \em is a subsystem \em of the dynamical system ${ \Phi}$ of ${\mathcal N}$, up to the continuous change $\psi $ of the state variables.

   From Definition \ref{definitionRicherNetwork} it is immediately deduced the following assertion:

    \em ${\mathcal N}' \sqsubseteq {\mathcal N}$ and ${\mathcal N} \sqsubseteq {\mathcal N}'$ if and only if their respective dynamical systems $\Phi'$ and $\Phi$ are   topologically conjugated. \em

     This means that the dynamics of ${\mathcal N}$ and ${\mathcal N}'$ are \lq\lq the same topological dynamical system,\rq\rq \   up to an homeomorphic change in their variables  which is called \em a conjugacy\em. So, we   deduce:

      \lq\lq $\sqsubseteq$\rq\rq \  \em is a partial order in the class ${\mathcal C}$ of synaptically equivalent networks up to conjugacies. \em

\end{definition}

 As an example, assume that the numbers $N$ and $N'$ of neurons of ${\mathcal N}$ and ${\mathcal N'}$ satisfy $N  = 2 N'$. Then $X_0= (x_1(0), x_2(0), \ldots , x_{2N'}(0))$, $X_0' = (x'_1(0), x'_2(0), \ldots, x'_{N'}(0))$. Define the function $\psi$ by:
 $$\psi(X_0') = \psi (x'_1, x'_2, \ldots, x'_{N'}) = $$ $$ = (x'_1, x'_1, x'_2, x'_2, \ldots, x'_{N'}, x'_{N'}) = X_0.$$
 If this function $\psi$ satisfies Equality (\ref{eqn006}), then each orbit $\{\Phi'(X'_0, t)\}_{t \geq 0}  $ of the dynamical system of ${\mathcal N}'$, is identified with one orbit $\Phi(X_0, t)$ of the dynamics of ${\mathcal N}$. Along this orbit $\Phi(X_0, t)$, each two consecutive identical neurons  have the same initial states, and thus also have coincident instantaneous states for all $t \geq 0$. Nevertheless, the whole dynamics $\Phi$ of the   network ${\mathcal N}$ also includes   many other different orbits, which are obtained   if the initial states of some pair  of consecutive identical neurons of ${\mathcal N}$ are mutually different.

  \begin{remark}
   \em From Definition \ref{definitionRicherNetwork}, since $\psi: M' \mapsto M$ is continuous and one-to-one, we deduce that the image   $\psi(M')$ is a submanifold of $M$ which is homeomorphic to $M'$. This is a direct application of  the \lq\lq Domain Invariance Theorem\rq\rq \  (see for instance  \cite{DomInvTheo}). Therefore:

   \vspace{-.4cm}

    $$\mbox{dim}(\psi(M')) = \mbox{dim}(M').$$

   \vspace{-.1cm}

    \noindent Besides, $\psi(M') \subset M$. So, $M$ contains the submanifold $\psi(M')$ that has the same dimension than $M'$. We deduce the following statement:

   \vspace{-.4cm}

\begin{equation}\label{eqn012}     {\mathcal N}' \sqsubseteq {\mathcal N} \ \Rightarrow \ \mbox{dim}(M') \leq \mbox{dim}(M).   \end{equation}

   \vspace{-.2cm}

    \noindent In extensum:

    \em If the dynamics of ${\mathcal N}$ is richer than the dynamics of ${\mathcal N}'$, then the dimension of the manifold $M$ where the dynamics of ${\mathcal N}$ evolves, is larger or equal than the dimension of the manifold $M'$ where the dynamics of ${\mathcal N}'$ evolves. \em
    \end{remark}

\vspace{-.3cm}

 From the above remark  we deduce the following  result:

 \vspace{-.2cm}

 \begin{proposition} {\bf  (Intermediate result in the proof of Main Theorems \ref{theoremDalePrinciple} and \ref{theoremCounterexample})}

 \label{propositionNumberOfNeuronsLarger}

   If ${\mathcal N}$ and ${\mathcal N}'$ are synaptically equivalent and if ${\mathcal N}' \sqsubseteq {\mathcal N}$, then the number of neurons of ${\mathcal N}$ is larger or equal than the number of neurons of ${\mathcal N'}$.
 \end{proposition}
  {\em Proof: } Both networks are synaptically equivalent; so, each neuron $i$ of ${\mathcal N}$ is structurally identical to some neuron (which we still call $i$) of ${\mathcal N}'$. This implies that the finite dimension of the   variable $x_i \in M_i$ in the network ${\mathcal N}$ is equal to the finite dimension of the corresponding   variable $x'_i \in M'_i$ in the network ${\mathcal N}'$. Thus,

   \vspace{-.5cm}

   \begin{equation}
   \label{eqn013}
   \mbox{dim}(M_i) = \mbox{dim}(M_i).\end{equation}

   \vspace{-.1cm}

   \noindent After Equality (\ref{eqn12}) applied to the networks ${\mathcal N}$ and ${\mathcal N'}$ respectively, we obtain:

   \vspace{-.4cm}

   \begin{equation}
   \label{eqn014}\mbox{dim}(M) = \sum_{i= 1}^N \mbox{dim}(M_i), \end{equation}

   \vspace{-.4cm}

   \begin{equation}
   \label{eqn015} \mbox{dim}(M') = \sum_{i= 1}^{N'} \mbox{dim}(M'_i),\end{equation}

   \vspace{-.2cm}

   \noindent where $N$ and $N'$ are the number of neurons of ${\mathcal N}$ and ${\mathcal N}'$ respectively.
   From Inequality (\ref{eqn012}) we have:

   \vspace{-.4cm}

   $$\mbox{dim}(M') \leq \mbox{dim}(M),$$

   \vspace{-.2cm}

   \noindent Finally, substituting  (\ref{eqn014}) and (\ref{eqn015}), we obtain:

   \vspace{-.4cm}

    $$\sum_{i= 1}^{N'}\mbox{dim}(M'_i) \leq \sum_{i= 1}^{N }\mbox{dim}(M_i),$$

   \vspace{-.2cm}

   \noindent and joining with    (\ref{eqn013})  we conclude $N' \leq N$, as wanted.

  \hfill $\Box$

\vspace{-.4cm}

\begin{definition}
{\bf (Dynamically  optimal networks)}
\label{definitionDynamicalOptimum}
\em We say that a network ${\mathcal N} \in {\mathcal C}$ is a \em dynamical  optimum conditioned to the synaptical  equivalence class ${\mathcal C}$  \em (i.e. within the class ${\mathcal C}$) if
   ${\mathcal N}' \sqsubseteq {\mathcal N}$ for all ${\mathcal N}' \in {\mathcal C}$.
\end{definition}

\begin{theorem}
\label{theoremExistenceofOptimalNetworks} {\bf (Existence of the dynamical optimum \\ -Intermediate result in the proof of Main Theorems \ref{theoremDalePrinciple} and \ref{theoremCounterexample})}

For any class ${\mathcal C}$ of synaptically equivalent neuronal networks there exists   a dynamical  optimum network conditioned to ${\mathcal C}$. This optimal network has the maximum number of cells among all the networks of the class ${\mathcal C}$.
\end{theorem}
{\em Proof: }
The class ${\mathcal C}$ of synaptically equivalent networks is characterized by the numbers $l$ and $s$ of homogeneous parts and synaptical units respectively, and by the real values $\Delta_{U, B}$ of the synaptical connections  between the dynamical units $U$ and the homogeneous classes $B \not \supset U$.

For each dynamical unit $U$, we denote by $ l_U \leq l-1$ the number of homogeneous classes $B \not \supset U$ such that $\Delta_{U, B} \neq 0$. Thus, $l_U \geq 1$ because each cell   $i \in U$ is not indifferent, and so, there exists at least one   nonzero synaptical action departing from  $i$.
(Recall that by Definitions \ref{definitionIdenticalCells} \ref{definitionHomogeneousPart}, the nonzero synaptical actions only exist between cells belonging to different homogeneous parts.)

    Construct a network ${\mathcal N}$ as follows:

    First, compose each dynamical unit $U$ with exactly $l_U$ cells.  Then,  there exists a surjective one-to-one correspondence $\Lambda_U$ between the set of cells  $i \in U$ and the set of homogeneous parts $B \not \supset U$ satisfying $\Delta_{U, B} \neq 0$.

      Second, define  the synaptical connections departing from each cell $i$ of each dynamical unit $U$, by the following equalities: \begin{equation}
\label{eqn007}
\Delta_{i, j} := \Delta_{U, B} \neq 0 \ \forall \ j \in B  \mbox{ if }   \ B = \Lambda_U(i),\end{equation}
\begin{equation}
\label{eqn007b}
\Delta_{i, j} := \Delta_{U, B} = 0 \ \forall \ j \in B  \mbox{ if } \ B \neq \Lambda_U(i).\end{equation}
   We will prove that the network ${\mathcal N}$ such constructed is dynamically optimal within the class ${\mathcal C}$:

 Fix any network ${\mathcal N}' \in {\mathcal C}$. Consider the dynamical systems $\Phi$ and $\Phi'$ corresponding to the networks ${\mathcal N}$ and ${\mathcal N'}$ respectively. Denote by $M$ and $M'$ the compact manifolds where $\Phi$ and $\Phi'$ respectively evolve. According to Definition \ref{definitionRicherNetwork}, to prove that ${\mathcal N}' \sqsubseteq {\mathcal N}$ it is enough to   construct a continuous one-to-one mapping $\psi: M' \mapsto M$ satisfying Equality (\ref{eqn006}).

 Let $X_0' \in M'$. For any cell $i' \in {\mathcal N}'$, the initial state $x'_{i'}(0)$ is a component of $X_0'$. Let us define $X_0 = \psi (X'_0) \in M$ satisfying Equality (\ref{eqn006}). To do so, we must define the initial state $x_{i}(0)$ of any cell $i$ of the network ${\mathcal N}$.

 So, {\bf fix $i \in {\mathcal N}$}. Denote by $U$ the synaptical unit to which $i$ belongs, and denote by $B = \Lambda_U(i)$ the unique homogeneous class of the network ${\mathcal N}$ satisfying  (\ref{eqn007}). We denote   \begin{equation}
 \label{eqn010a}
  \Delta_{i,B}  := \Delta_{i,j} = \Delta_{U, B}  \neq 0 \ \ \forall \ j \in B,\end{equation}
  where $ i\in U $ and $ B= \Lambda_U(i).$

 Since ${\mathcal N}'$ is synaptical equivalent to ${\mathcal N}$ (because both networks ${\mathcal N}$ and ${\mathcal N}'$ belong to the same class ${\mathcal C}$), we apply  Definition \ref{definitionEquivalentNetworks} to deduce the following equalities:
 \begin{equation}
 \label{eqn010b} \Delta_{U', B'} = \Delta_{U, B} = \Delta_{i,B} \neq 0, $$ $$ \mbox{ where } \ U'= \varphi (U), \ B'= \varphi (B).\end{equation}
 From Definition  \ref{definitionSynapticalUnit},   there exists a unique     cell $i' \in U'$  such that \begin{equation}
 \label{eqn010c}
 \Delta_{i', B'} = \Delta_{U, B'} \neq 0.\end{equation} Summarizing, for any fixed neuron $i \in U \subset {\mathcal N}$ we have constructed a unique cell $i' \in U' = \varphi (U) \subset {\mathcal N}'$ such that Equalities (\ref{eqn010a}), (\ref{eqn010b}) and (\ref{eqn010c}) hold. In other words, we have constructed a mapping $\Pi: \ i' = {\Pi  } (i)$, defined from the synaptical equivalence between the networks ${\mathcal N}$  and ${\mathcal N}'$,   such that:
 \begin{equation}
 \label{eqn009}
 \Delta_{\Pi(i), \varphi(B)} = \Delta_{i, B} \neq 0 \ \ \forall \ i \in {\mathcal N},\end{equation}
 where $B = \Lambda_U(i)$ is the unique homogeneous class in ${\mathcal N}$ satisfying   (\ref{eqn007}).

 \noindent {\bf Assertion A: } The mapping $\Pi$   transforms each cell  $i$ of the network ${\mathcal N}$ into the  cell $i'= \Pi(i)$ in the network ${\mathcal N}'$, \em which is structurally identical to $i$. \em

 In fact, assertion A follows from the fact that ${\mathcal N}$ and ${\mathcal N}'$ are synaptically equivalent (cf. Definition \ref{definitionEquivalentNetworks}) and from Equality (\ref{eqn009}).

 Let us prove that $\Pi$ is surjective. In fact, for each $i' \in {\mathcal N}'$, there exists at least one homogeneous part $B'$ such that $\Delta_{i', B'} \neq 0$, because $i'$ is not indifferent. By Definition \ref{definitionEquivalentNetworks}, $B' = \varphi(B)$ where $\varphi$ is a one-to-one and surjective transformation between the homogeneous parts  of ${\mathcal N}$ and ${\mathcal N'}$. Therefore, there exists a unique homogeneous part $B$ of ${\mathcal N}$ such that $\Delta_{i', B'} = \Delta_{U, B} \neq 0$, where $B= \varphi^{-1}(B')$, $U = \varphi^{-1}(U')$ and $i' \in U'$. By construction of the network ${\mathcal N}$,  if $\Delta_{U, B} \neq 0$, then there exists a unique $i \in U $ such that $\Delta_{U, B} = \Delta_{i, B}$. Then, we deduce that
  $\Delta_{{i'}, \varphi (B)} = \Delta_{i,B} \neq 0$. Joining with (\ref{eqn009}), and recalling that for  each synaptical unit $U'$ there exists at most one cell $i' \in U'$ such that $\Delta_{i', B'} \neq 0$, we deduce $i' = \Pi(i)$. This proves that $\Pi$ is surjective.

 We define the initial state $x_i(0)   $ of the   cell $i \in {\mathcal N}$ by $$x_i(0) := x_{\Pi(i)}(0),$$ and the mapping $\psi: M' \mapsto M$ by
 \begin{equation}
 \label{eqn008}
 \psi (X'_0) = X_0 \mbox{ such that }$$ $$ \ x_i(0) = x'_{\Pi(i)}(0) \ \ \forall \ i \in {\mathcal N}.\end{equation}
  The mapping $\psi$ is continuous because the components $x_i(0)$  of $\psi (X'_0)$ are components $x'_i(0)$ of $X'_0$. Thus, small increments in the components $x'_{i'}(0)$ of  $X'_0$ imply    small increments in the components $x_i(0)$ of $X_0 = \psi(X'_0)$. Besides, the mapping $\psi$ is one-to-one (but non necessarily surjective). In fact, if $X'_0  \neq Y'_0$ then, at least one component $x'_{i'}(0)$ of $X'_0$ differs from the respective component $y'_{i'}(0)$ of $Y'_0$. Since $\Pi$ is surjective,  there exists $i$ such that $i'= \Pi(i)$. So, applying Equality (\ref{eqn008}) we obtain $x_i(0) \neq y_i(0)$, where $y_i(0) = y'_{\Pi(i)}(0)$. Thus $\psi(X'_0)  \neq \psi(Y_0')$ proving that $\psi$ is one-to-one.

To end the proof of the first part of Theorem \ref{theoremExistenceofOptimalNetworks}, it is now enough   to check that the mapping $\psi$ satisfies Equality (\ref{eqn006}):

From Equality (\ref{eqn008}) and from the surjectiveness of $\Pi$, for each initial state $X'_0$ of the network ${\mathcal N}'$, and for each neuron $i' \in {\mathcal N}'$, the corresponding set of neurons $i \in \Pi^{-1}(i') \subset {\mathcal N}$ have  initial states  $x_i(0)$ which equal    $x'_{i'}(0)$. Besides, from Assertion A, $i \in \Pi^{-1}(i')$ and $i'$ are structurally identical. Now, we consider Equalities (\ref{eqn010a}), (\ref{eqn010b}) and (\ref{eqn009}), applied to any neuron $j \in {\mathcal N}$ and $j' = \Pi(j) \in {\mathcal N}'$, in the respective roles of $i$ and $i'= \Pi (i)$. We deduce that the synaptical interaction  jumps $\Delta_{j', i'}$ that $i'$ receives from any other neuron $j' \in {\mathcal N}'$ coincides with the synaptical interaction jumps $\Delta_{j, i}$ that $i \in \Pi^{-1}(i')$ receives from $ j \in \Pi^{-1}(j')$ in the network ${\mathcal N}$. Therefore,   both $i'$ and $\Pi^{-1}(i')$ satisfy the same impulsive differential equation (\ref{eqn005}). Besides, their respective initial conditions $x'_{i'}(0)$ and $x_{\Pi^{-1}(i')} (0)$ coincide, due to Equality (\ref{eqn008}). Since the solution of the impulsive differential equation (\ref{eqn005}) that satisfies a specified initial condition is unique, we deduce the following statement:

 \em For any instant $t\geq 0$ the    state $  x_{i'}(t)$ coincides with the instantaneous state $ x_{i}(t)$, where $i \in \Pi^{-1}(i')$. \em

 Recalling Definition \ref{definitionDynamicsofN} of the dynamics $\Phi$ and $\Phi'$ of the networks ${\mathcal N}$ and ${\mathcal N}'$ respectively, we deduce: $$X(t) = (  \ldots, x_i(t), \ldots) =: \Phi(X(0), t), $$ $$ X'(t) = ( \ldots, x'_{\Pi(i)}(t), \ldots) =: \Phi'(X'(0), t).$$
 Applying again Equality (\ref{eqn008}), which defines the mapping $\psi$ for each fixed instant $t \geq 0$ as the new initial state, we conclude
 $$\Phi(\psi (X'_0), t) = \psi (\Phi'(X'_0), t),$$
 proving Equality (\ref{eqn006}), as wanted.

 We have proved that ${\mathcal N}' \sqsubseteq {\mathcal N}$ for all ${\mathcal N}' \in {\mathcal C}$. Thus, in each synaptical equivalence class ${\mathcal C}$ there exists a network ${\mathcal N}$ that is the dynamical  optimum conditioned to ${\mathcal C}$.

 Now, let us prove the second part of   Theorem \ref{theoremExistenceofOptimalNetworks}. We have   to show that   the number $N$ of neurons in ${\mathcal N}$ is the maximum   number  of neurons of all the networks in the class ${\mathcal C}$. In fact, since ${\mathcal N}' \sqsubseteq {\mathcal N}$, after Proposition \ref{propositionNumberOfNeuronsLarger} we get $N' \leq N$, where $N'$ is the number of neurons of ${\mathcal N}'$, for all ${\mathcal N}' \in {\mathcal C}$.
\hfill $\Box$

\vspace{-.5cm}

\section{End of the proof of Dale's Principle} \label{sectionDalePrinciple}

\vspace{-.3cm}

Let $\aleph$ be the set of all the neuronal networks according to the mathematical model defined in Section \em \ref{sectionModel}\em.
Let ${\mathcal C} \subset \aleph$ be a fixed class of synaptically equivalent  networks.
The purpose of this section is to end the proof of the following Main  Theorem of the paper:

%\newpage

\begin{theorem}
\label{theoremDalePrinciple} {\bf (Dale's Principle is necessary for the dynamical optimization)}

 If  ${\mathcal N}$ is the dynamical   optimum network conditioned to ${\mathcal C}$,
then  all the neurons of ${\mathcal N}$ satisfy Dale's Principle.

\em Namely, any neuron of   ${\mathcal N}$ is either inhibitory or excitatory.
\end{theorem}

\vspace{-.3cm}

 \noindent {\em End of the proof of Theorem }\ref{theoremDalePrinciple}: Let ${\mathcal N}$ be the dynamical optimum among the networks in ${\mathcal C}$.  Therefore,

 \vspace{-.3cm}

  $${\mathcal N}' \sqsubseteq {\mathcal N} \ \ \forall \ {\mathcal N}' \in {\mathcal C}.$$
  Thus, applying Proposition \ref{propositionNumberOfNeuronsLarger}, the numbers $N$ and $N'$ of neurons in ${\mathcal N}$ and ${\mathcal N}'$ respectively, satisfy

 \vspace{-.3cm}

  \begin{equation}
  \label{eqn011} N' \leq N.
  \end{equation}

 \vspace{-.2cm}

  \noindent Denote by $\Delta_{j,h}$ the synaptical  action  from the neuron $j$ to the neuron $h \neq j$ in ${\mathcal N}$, for any $j \in \{1, \ldots, N\}.$ {\bf Assume by contradiction that there exists a neuron $ i \in  {\mathcal N}$ which is mixed}, according to Definition \ref{definitionExcitInhibitMixed}. Let us fix such a value of $i$. Now, we construct a new  network ${\mathcal N}' \in {\mathcal C}$ as follows:

  First, include in ${\mathcal N}'$ all the   neurons $j \in {\mathcal N}$, in particular $j= i$. Define  in ${\mathcal N}'$  the synaptical interactions $\Delta'_{j,h}$  as follows:
  $$\Delta'_{j,h} := \Delta_{j,h} \ \ \forall \ h \neq j \mbox{ if } j \neq i.$$
  $$\Delta'_{i,h} := \max\{0,\Delta_{i,h}\} \ \ \forall \ h \neq i.$$

  Second, add one more neuron in ${\mathcal N}'$, say the $N+1$-th. neuron, which we make, by construction, structurally identical to the $i-$th. neuron.
  Define
  $$\Delta'_{N+1, h} := \min\{  \Delta_{i,h}, 0\} \ $$ $$\ \forall \   \ h \in \{1, 2, \ldots, N\} \mbox{ such that } h \neq i,$$
  $$\Delta'_{N+1, i'} := 0, \ \ \Delta_{i', N+1}:= 0.$$
  The new neuron $N+1$ is not indifferent in the network ${\mathcal N}'$ because $i$ is mixed in ${\mathcal N}$. (So, there exists $h \neq i$ such that $\Delta_{i,h} < 0$.)

  It is immediate to check that ${\mathcal N}'$ is synaptically equivalent to ${\mathcal N}$. In fact, all the    neurons, except the added one $N+1$-th. cell in ${\mathcal N}'$, are respectively structurally identical in the networks ${\mathcal N}$ and ${\mathcal N}'$. Besides, all the synaptical interactions, except those that depart from   $i$ and $N+1$, are the same in both networks. Finally, also the nonzero synaptical interactions that depart from $i$ in the network ${\mathcal N}$, are equal, either to the synaptical interactions that depart from $i$ in ${\mathcal N}'$ (if positive), or to those that depart from the new neuron $N+1$ in ${\mathcal N}'$ (if negative). So, ${\mathcal N}'$ is synaptically equivalent to ${\mathcal N}$. In other words, ${\mathcal N}' \in {\mathcal C}$. To end the proof, we note that  the number $N'$ of neurons of ${\mathcal N}'$ is $N' = N+1 > N$, contradicting Inequality (\ref{eqn011}). \hfill $\Box$

  \vspace{-.5cm}

 \section{Counter example} \label{sectionCounterexample}

 \vspace{-.3cm}

 The purpose of this section is to exhibit a counter-example  that shows that the converse of Main Theorem \ref{theoremDalePrinciple} is false (Theorem \ref{theoremCounterexample}).

 Theorem \ref{theoremCounterexample} is the second Main Theorem of the paper. Its   proof    is deduced from the intermediate results that were previously obtained along the paper, and  is ended by   showing the explicit counter-example from Figures \ref{Figure1}, \ref{Figure2} and \ref{Figure3}.

 \vspace{-.3cm}

 \begin{theorem}
 \label{theoremCounterexample} {\bf (Dale's Principle is not sufficient for the dynamical optimization)}

 There exist    neuronal networks according to the mathematical model of Section \em \ref{sectionModel} \em that satisfy Dale's Principle and are not dynamically optimal conditioned to their respective synaptical equivalence classes.
 \end{theorem}

\vspace{-.3cm}

 \noindent {\em End of the proof of Theorem} \ref{theoremCounterexample}:
 We will show an explicit  example of a dynamical \em suboptimal \em network ${\mathcal N}$ within a synaptical equivalence class ${\mathcal C}$, such that ${\mathcal N}$ \em satisfies \em Dale's Principle. We will exhibit such an example with $N= 5$ neurons, but it can be repeated (after obvious adaptations) with any arbitrarily chosen number $N \geq 3$.

 Consider the network ${\mathcal N}$ of Figure \ref{Figure1}. Assume, for instance, the following signs for the nonzero synaptical interactions:

 \noindent $\Delta_{4,i} >0$ for   $i= 1,2,3$,

 \noindent $\Delta_{5,i} < 0$ for $i = 1,2,3, 4$,

 \noindent $\Delta_{1,4} < 0$, $\Delta_{2,4} >0$, $\Delta_{3,5} < 0 $.

 Then, the neurons $2$ and $4$ are excitatory and the neurons $1,3$ and $5$ are inhibitory. Thus, all the neurons of  the network ${\mathcal N}$ satisfy Dale's Principle.

 As shown in Section \ref{sectionEquivalentNetworks}, the network ${\mathcal N}'$ of Figure \ref{Figure3} is synaptically equivalent to the network ${\mathcal N}$ of Figure \ref{Figure1}. In other words, both networks ${\mathcal N}$ and ${\mathcal N}'$ belong to the same equivalence class ${\mathcal C}$. Since ${\mathcal N}'$ has exactly 6 neurons and ${\mathcal N}$ has 5 neurons, applying Proposition \ref{propositionNumberOfNeuronsLarger} we deduce that ${\mathcal N}' \not \sqsubseteq {\mathcal N}$. Thus ${\mathcal N}$ is not the   optimal network of its class ${\mathcal C}$.
 \hfill $\Box$

 \vspace{-.4cm}

 \section{Final Comments} \label{sectionConclusions}

 \vspace{-.3cm}

 In Section \ref{sectionModel} we posed the mathematical  simplified (but general) model of biological neuronal networks, by a  system (\ref{eqn005})  of deterministic impulsive differential equations. In its essence, this model was taken from \cite{Izhikevich_PulseCoupledDynamicsExcitabilityBursting} (some particular conditions of the model were also taken from \cite{SpikingNeuronModels,ErmentroutDynamicsModelsExcitability,
 ImpulsiveDifferentialEquationsSpikingSynapsesModels,CatGuiraud,Coombes2013} and from the bibliography therein).

 On the one hand, the mathematical model is an idealized simplification of the network, because the spiking of each neuron  is reduced to an instantaneous reset, without delay, of its membrane potential. Also  the synaptical actions are   assumed to be instantaneous and have no delay.

 On the other hand, the abstract mathematical model is  general, since we   require neither particular formulae, nor numerical specification, nor computational algorithms  for the functions $f_i$, $F_1$ and $\Delta_{i,j}$ of   Equations (\ref{eqn001}), (\ref{eqn003}) and (\ref{eqn005}), nor    specific values  for the parameters.

 In Section \ref{sectionGraph&Layers} we defined the homogeneous parts of the network, composed by mutually identical cells. The groups of neurons, which we call \lq\lq synaptical units\rq\rq, are formed by   structurally identical and synaptically representative neurons. In Proposition \ref{propositionUnits} we proved that any neuronal network, according to the mathematical model described in Section \ref{sectionModel}, is decomposable in more than one homogeneous part, and that each homogeneous part is decomposable into pairwise disjoint dynamical units. Then, a simplified graph, which we called \lq\lq inter-units graph\rq\rq \ mathematically represents the statical structure of the synaptical connections among the groups of neurons in the network. This theoretical approach have rough similitudes with empirical research in Neuroscience \cite{DistributionInhibitoryExcitatotySynapsesOnHippocampalCells}, for which the structure of synaptical connections among groups  of neurons or regions in the brain is studied, regardless how many neurons exactly compose each region.

 In Section \ref{sectionEquivalentNetworks} we conditioned the study to a fixed family of  networks that are mutually synaptically equivalent. We denote this family by ${\mathcal C}$, and call it a class. Even if this condition may appear as a restriction, it is not. In fact, first, any neuronal network (provided that it is mathematically modelled by the equations of Section  \ref{sectionModel}), belongs to one such a    class ${\mathcal C}$. Second,   all the results that we proved along the paper stand for any arbitrarily chosen class ${\mathcal C}$ of synpatically equivalent networks.

 Each class ${\mathcal C}$ of   synaptically equivalent networks  gives a particular specification for the number  of synaptical units  and for the inter-units graph. This specification implies   a particular statical \lq\lq anatomy\rq\rq \ in the synaptical structure of the network, described by the different groups of mutually identical neurons (and not by the neurons themselves). Each group of neurons is a synaptical unit that has a characteristic functional role in the complex    synaptical structure of the network.

 Roughly speaking, a class ${\mathcal C}$ of mutually synaptically equivalent neuronal networks works as an abstraction of the   following analogous example:  When a Neuroscientist studies the nervous system of certain species of animals, he is investigating  \em a  class of neuronal networks \em  composed by a relatively large amount of  particular cases that are indeed different networks (one particular case for each individual of the same species). But all the neuronal networks in that class share a certain structure, which is given, for instance, by the genetic neurological characteristics of the species. Some   type of synaptical connections between particular \em groups \em of neurons with specific physiological roles, is shared by    all the healthy individuals of the species. However, the exact number of neurons, and the exact number and weight of synaptical connections between particular neurons,  may  vary from one individual to another of the same species, or from   an early age to a mature age of the same individual.

 In Section \ref{sectionDynamicalOptimum} we studied the abstract dynamical system of any neuronal network defined by the mathematical model of Section \ref{sectionModel}, and conditioned to a certain fixed class ${\mathcal C}$ of mutually synaptically equivalent networks. In Theorem \ref{theoremExistenceofOptimalNetworks} we proved that (theoretically) a dynamically optimal network exists in each class ${\mathcal C}$.

 The proof of Theorem \ref{theoremExistenceofOptimalNetworks} is   constructive: first, we defined a particular network ${\mathcal N} \in {\mathcal C}$, and second, we proved that ${\mathcal N}$ is the richest network of its class. This means that ${\mathcal N}$ would  potentially exhibit the most diverse set of dynamical responses (orbits in the future)  when  external  signals change the instantaneous state of some of its neurons.

 Since the system is assumed to be deterministic, any network according to this model will reproduce a unique response if the same instantaneous state occurs for   \em all \em its neurons. So, the space of responses is represented by the space of instantaneous states (or initial states, if time $T$ is translated to become 0).  Nevertheless, this space may change from one network to another of the same synaptical equivalence class ${\mathcal C}$. If we assumed  that the \lq\lq natural pursued  aim\rq\rq \ in the development of a biological neuronal network were to optimize the space of dynamical responses under stimulus, preserving the same   characteristic and functional structure between groups of cells, then, theoretically, the final  (but maybe never arrived)   network would be ${\mathcal N}$, constructed in the proof of Theorem \ref{theoremExistenceofOptimalNetworks}.

 In Section \ref{sectionDalePrinciple} we proved Theorem \ref{theoremDalePrinciple}, which is one of the main results of the paper. It states that the dynamically optimal network ${\mathcal N} $ in the class ${\mathcal C}$ must satisfy Dale's Principle (i.e. all its neurons are either excitatory or inhibitory but not mixed). So, if the natural pursued aim in the development of the neuronal network were to optimize the space of possible dynamical responses, then the tendency of the network during its plastic phases will provoke that as many neurons as possible satisfy Dale's Principle. From this point of view, Theorem \ref{theoremDalePrinciple} shows that Dale's Principle is a consequence of an optimization process. So, it gives a mathematically possible   answer to the following epistemological question:

 \em Why does Dale's Principle hold  for most neurons of most biological neuronal networks? \em

 Mathematical answer: Because maybe biological  networks evolve  pursuing the theoretical optimum or richest dynamics,  conditioned to preserve     the synaptical connections among its different homogeneous groups of neurons.

 Finally, in Theorem \ref{theoremCounterexample} we proved that Dale's Principle is not enough for the neuronal network be a dynamical optimum within its synaptically equivalence class. In other words, Dale's Principle would be just a stage of a plastic optimization process of the neuronal network, but its validity does not ensure that the end of that hypothetical process of optimization has been arrived.

 \vspace{.2cm}

 \noindent {\bf Acknowledgements}

 \noindent The author thanks the anonymous referees for their valuable comments and suggestions,   Agencia Nacional de Investigaci\'{o}n e Innovaci\'{o}n (ANII)  and Comisi\'{o}n Sectorial de Investigaci\'{o}n Cient\'{\i}fica (CSIC) of the Universidad de la Rep\'{u}blica, both institutions of Uruguay.

%%%%%%%%%%%%%%%%%%%%%%%%%%%%
%%%%%%%%%%%%%%%%%%%%%%%%%%%%%

%\section{T\'{\i}tulo}

%\vspace{1cm}

%\vfill
%
%\noindent \texttt{Revised version: June 27th 2013. }
%
%\noindent \texttt{Journal "Applied Mathematics" \\ Accepted ID 7401671 \\ www.scirp.org/journal/am}
%
%\noindent \texttt{ISSN: 2152-7393 \\ Special issue  "Biomathematics"}

\end{document}